\documentclass[twocolumn]{aastex63}

\usepackage{commath}

\submitjournal{ApJL}

\shorttitle{Pop II star cluster }
\shortauthors{Latif \& Schleicher}

\begin{document}

\title{Formation of Pop II star clusters  in the aftermath of a pair instability supernova}



\correspondingauthor{Muhammad A. Latif}
\email{latifne@gmail.com}

\author{Muhammad A. Latif}
\affiliation{Physics Department, College of Science, United Arab Emirates University, PO Box 15551, Al-Ain, UAE}

\author{Dominik Schleicher}
\affiliation{Astronomy Department, Universidad de Concepci\'on, Barrio Universitario, Concepci\'on, Chile}

%
\begin{abstract}
Pop II stars formed a few hundred million years after the Big Bang were key drivers of cosmic reionization and  building blocks of high redshift galaxies.  How and when  these stars formed is a subject of ongoing research.  We conduct cosmological radiation hydrodynamical  simulations to  investigate the formation of Pop II star clusters in  dark matter halos forming at $z=10-25$ in the aftermath of a pair instability supernova (PISN). Our simulations model the formation of Pop III and Pop II stars in a self-consistent manner along with their radiative, chemical and SN feedback in halos of   $\rm 5 \times 10^5- 7 \times 10^7~M_{\odot}$. We find that a PISN evacuates the gas from halos  $\rm \leq 3  \times 10^{6}~M_{\odot}$  and  thereafter shuts off in situ star formation for  at-least 30 Myr.  Pop II stellar clusters of $\rm  923~M_{\odot}$  and $\rm 6800~M_{\odot}$ form in halos of $\rm 3.8 \times 10^7~M_{\odot}$ and $\rm 9 \times 10^7~M_{\odot}$, respectively.   The mode of star formation is highly episodic and  mainly regulated by Pop II SN feedback. The average star formation rates are  $\rm 10^{-5}-10^{-4}~M_{\odot}/yr$ and the star formation efficiency is less than   1\%. 

\end{abstract}


\section{Introduction} \label{sec:intro}

High redshift surveys have observed more than eight hundred galaxies within the first billion years after the Big Bang including candidate galaxies up to $z \sim 11$ \citep{Bouwens16,Oesch16,Lam19,Bowler20}. These surveys  have shifted the  observational frontier up to the cosmic dawn.  High redshift galaxies observed at $z\sim 10$ are potential hosts of the first stellar populations, key drivers of cosmic reionization and  metal enrichment in the universe.   The questions of how and when these galaxies formed stars have stimulated a lot of theoretical interest and the upcoming James Webb Space Telescope  is  expected to further unveil properties of their stellar populations.

The first stars known as Population III (Pop III) stars are presumed to be formed in  dark matter halos of $\rm 10^5-10^6~M_{\odot}$ at $z=20-30$.  Baryonic collapse in these minihalos is triggered by molecular hydrogen which can cool the gas down to 200 K in the absence of metals and lead to the formation of Pop III stars.  The numerical simulations of Pop III stars suggest  a wide range of possible masses from $\rm 1-1000~M_{\odot}$  \citep{Abel2000,Bromm2002,Clark11,Latif13ApJ,Stacy16,Riaz18,Sugimura20}.  These metal free stars are hotter than the present day stars, produce copious amounts of radiation, and photo-ionize the gas clouds in the surrounding medium \citep{Schaerer02,Whalen08,Whalen2013b}. They either go off   as a  PISN for a stellar mass between $\rm 140-260~M_{\odot}$, or type II  SN if the mass ranges from $\rm 11-40 ~M_{\odot}$. For masses above $\rm 260~M_{\odot}$ they directly collapse into a black hole.


In the aftermath of SNe,  metal lines and dust grains  cool the gas down to the CMB temperature  above  a critical metallicity of $\rm  \sim 3 \times 10^{-4}~Z_{\odot}$, leading to the formation of Pop II stars  \citep{Schneider03,Omukai05,Glover07,Wise2012a,Bovino14,Latif2016dust}.  The process of metal enrichment  is highly inhomogeneous  \citep{Chen17, Hartwig19} and can be external through SN winds \citep{Smith15} or via fallback of metal rich gas \citep{Ritter15,Chiaki19}. Various studies have  explored the impact of dust  cooling suggesting that it can operate at an even lower metallicity of $\rm \sim 10^{-5}~ Z_{\odot} $ \citep{Dopcke11, Dopcke13,Bovino16}.  \cite{SS14} performed simulations employing sink particles  and evolved one of the  clumps  for 7000 years at a fixed metallicity of $ \rm 10^{-2}~Z_{\odot}$. They found that  a star cluster of subsolar up to a few solar masses is formed. \cite{Smith15} explored the impact of dust cooling  in an externally metal enriched halo and  recently \cite{Chiaki19}  investigated the impact of fallback from a core collapse SN by modeling dust/metal  yields. These studies found that rapid dust cooling at high densities can lead to the formation of metal poor stars. 

While metal enrichment and metal poor star formation have been explored to some degree, the important question if and how a Pop II star cluster may form after a PISN  so far has not been answered. PISNe are about a hundred times more powerful than type II SNe and have much higher metal yields.  They may unbind low mass halos and  impact  the gas dynamics.  It is not clear how massive Pop II stellar clusters have formed in the first atomic cooling halos and what their properties have been. 

In this study we explore the formation of Pop II stellar clusters in the aftermath of a PISN. We self-consistently model the formation of Pop III and Pop II stars along with their radiative, chemical and mechanical feedback in cosmological simulations. We employ the radiative transfer module MORAY coupled with hydrodynamics to model UV feedback  from each Pop III and Pop II star particle. In total, we perform  five cosmological radiation hydrodynamical simulations of halos with mass ranging from  $\rm 5 \times 10^5- 7 \times 10^7~M_{\odot}$ at a maximum physical resolution of  2000 AU. Our simulations follow the evolution for about 80 Myr after the formation of first Pop III star and provide estimates of Pop II cluster masses in these halos. In section 2, we present our simulation setup, we present our findings in section 3 and discuss our conclusions in section 4.

\section{Numerical Method} \label{sec:methods}

Simulations are carried out using the cosmological hydrodynamics  code Enzo \citep{Enzo14} coupled with the radiative transfer module MORAY \citep{Wise11} to model radiative feedback from stars. We employ the MUSIC package \citep{Hahn11} to generate cosmological initial conditions at $z=150$ with a root grid resolution of $\rm 256^3$ and further employ two nested grid refinement levels in a computational periodic box of  size 1 Mpc/h. We make use of  the \textit{must refine particle} approach  to add refinement in the Lagrange volume of  two times the viral radius of the halo. Each DM particle is further split into 13 daughter particles  in the region of interest  and this approach yields an effective DM resolution of $\rm \sim 5~M_{\odot}/h$.  We further employ 13 additional  refinement levels during  the course of the simulations which results in a physical spatial resolution of about 2000 AU.  Our refinement criteria are based on particle mass resolution, the baryonic over-density and the Jeans refinement of at-least 16 cells \citep{Latif19}.

In total, we have simulated five halos of  $\rm 5 \times 10^5-9 \times 10^7~M_{\odot}$, at $z=26, ~ z=20, ~z=15,~z=13.6, ~z=12.5$, respectively. We turn on star formation in the halo soon after reaching the maximum refinement level and simultaneously  switch on both radiative and SN feedback.  Our recipes for star formation and stellar feedback are based on \cite{Wise08A}  \& \cite{Wise12} and  similar to  \cite{Latif18} \& \cite{Latif20b}.  A Pop III star particle is created when a cell meets the following criteria; 1)  an over-density of $5 \times 10^5$ ($\rm 10^3 ~cm^{-3}$ at $z=10$, II) an  $\rm H_2$  fraction of $ \rm \geq 5 \times 10^{-4}$, III) convergent flow.  The requirement of a minimum  $\rm H_2$ fraction ensures that Pop III stars form in molecular clouds. Each Pop III star particle represents a single star whose mass is randomly sampled from the Salpeter type initial mass function with mass range from $\rm 1-300~M_{\odot}$. In our simulation, we are unable to resolve an individual Pop II star due to the computational constraints and therefore, a single Pop II star particle represents a small cluster of stars.  Our criteria for Pop II stars are similar to the ones for Pop III without the requirement of a minimum  $\rm H_2$ fraction and they are distinguished based on the metallicity. Pop II stars are allowed to form in cells with $\rm T  < 1000 ~K$ and a minimum metallicity of $\rm 10^{-4}~Z_{\odot} $. 

The radiative feedback from stars (both Pop III and Pop II) is modeled using the ray tracing module MORAY  \citep{Wise11}  self-consistently linked with hydrodynamics. Pop III and Pop II stars are considered monochromatic sources of radiation with photon energies of 29.6 eV and 21.6 eV, respectively. For Pop III stars, we take the mass-dependent ionizing and Lyman Werner luminosities from \cite{Schaerer02} while Pop II stars emit $\rm 2.4 \times 10^{47} photons/s/M_{\odot}$ \citep{Schaerer2003}.  For SN feedback from Pop III stars, we  consider both PISN and type II SN for stellar masses between $\rm 140-260~M_{\odot}$ and $\rm 11-40~M_{\odot}$, respectively.  Pop II stars generate $\rm 6.8 \times 10^{48} erg/s/M_{\odot}$ after 4 Myr. The SN energy for both Pop III and Pop II stars is distributed in the surrounding $\rm 3^3$ cells and all stars forming in a sphere of 1 pc are merged to reduce the amount of ray tracing. Pop II stars live for 20 Myr corresponding to the lifetime of an OB star. For further details see \cite{Wise12}.  Our chemical model solves the non-equilibrium time dependent rate equations of primordial species ($ \rm H, ~H^+,~ H^-, ~He,~ He^+, ~He^{++},~ H_2, ~H_2^+, ~e^-$) based on \cite{Abel97} and is coupled with MORAY. It includes various cooling and heating processes for primordial chemistry, metallicity dependent metal line cooling from \cite{Glover07} in the temperature regime of $\rm 100-10^4~K$ and above $\rm 10^4~K$ cooling from \cite{Sutherland93}.  We consider a background flux of  strength unity in units of $\rm J_{21}=10^{-21}~erg/cm^2/s/Hz/Sr$.

\section{Results} \label{sec:results}

In total, we have simulated  five halos of  masses  $\rm 9 \times 10^7~M_{\odot}$, $\rm 3.8 \times 10^7~M_{\odot}$, $\rm 5 \times 10^5~M_{\odot}$, $\rm 1 \times 10^6~M_{\odot}$, $\rm 3.0 \times 10^6~M_{\odot}$,  named as halo 1, halo 2, halo 3, halo 4, halo 5, respectively. We follow the gravitational collapse in these halos until the central gas cloud is sufficiently cooled via $\rm H_2$ and collapsed to the densities  of $\rm 10^{-18}~g/cm^{3}$. Star formation in the halo is activated at this stage and a Pop III star of $\rm 182 ~M_{\odot}$ forms at the center of each halo.  The radiative feedback from the Pop III star is modeled with the 3D radiation transport algorithm MORAY coupled to hydrodynamics  for about 2.2 Myr corresponding to its lifetime. Radiation from the  star photo-ionizes the surrounding gas and  photo-dissociates $\rm H_2$. A HII region develops around the star and the density drops down to $\rm 10^{-25}~g/cm^{3}$.  At the end of its life, the star dies as PISN and deposits $\rm E_{PISN}= 3.6 \times 10^{52} ~erg$ into the halo.  Consequently, halos 3, 4 \& 5 (below $\rm 10^7~M_{\odot}$) are evaporated as the energy deposition from PISN exceeds their binding energy. Subsequently, we evolved these halos and found that star formation shuts off  for at-least 31 Myr.  Therefore, we hereafter  discuss star formation in  the rest of the two halos with masses larger than  $\rm 10^7~M_{\odot}$.

\begin{figure*}
\hspace{-6.0cm}
\centering
\begin{tabular}{c c}
\begin{minipage}{6cm}
\vspace{-0.2cm}
\hspace{-2.0cm}
\includegraphics[scale=0.9]{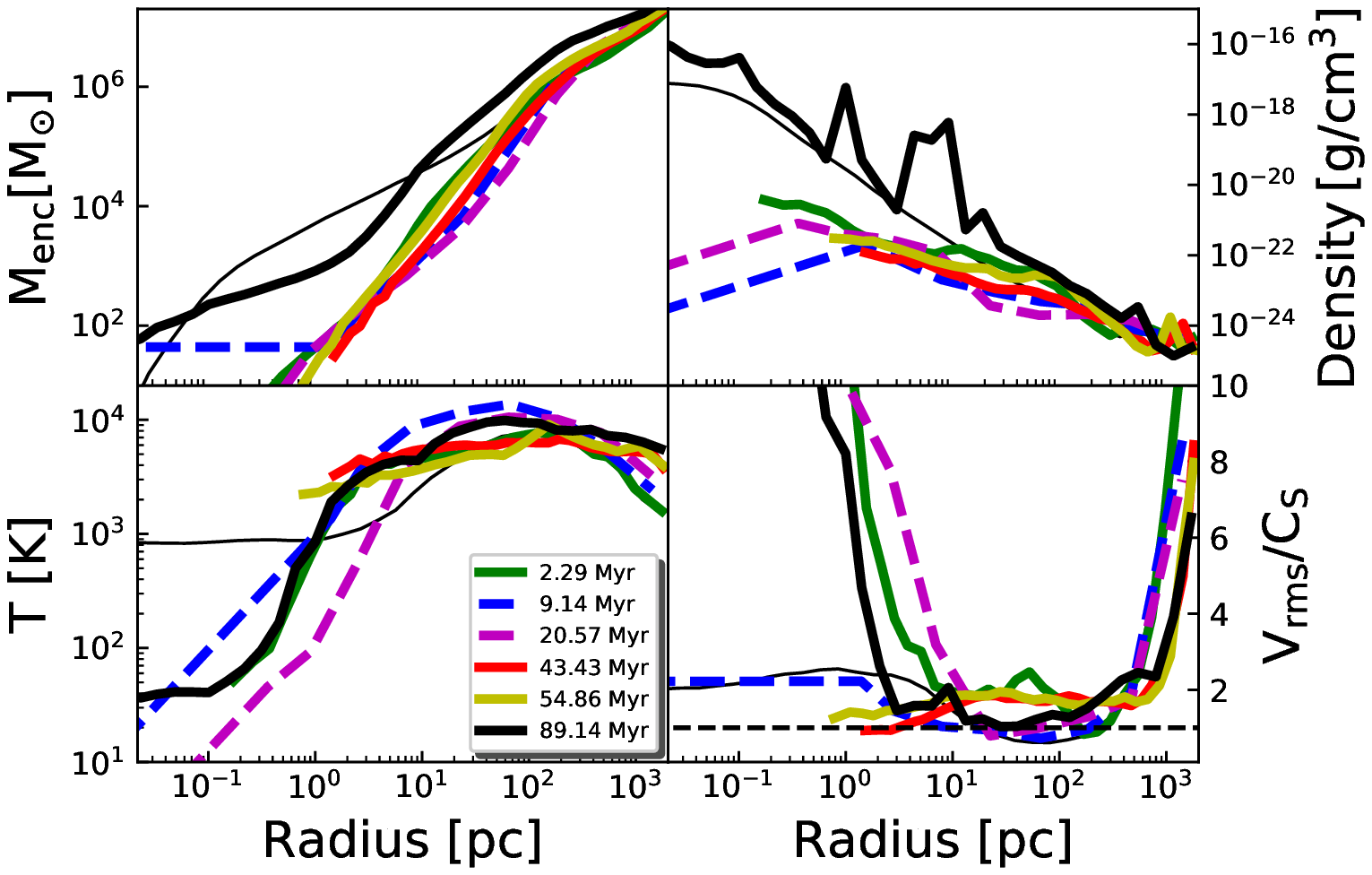}
\end{minipage} \\
\begin{minipage}{6cm}
\vspace{-0.5cm}
\hspace{-2.0cm}
\includegraphics[scale=0.9]{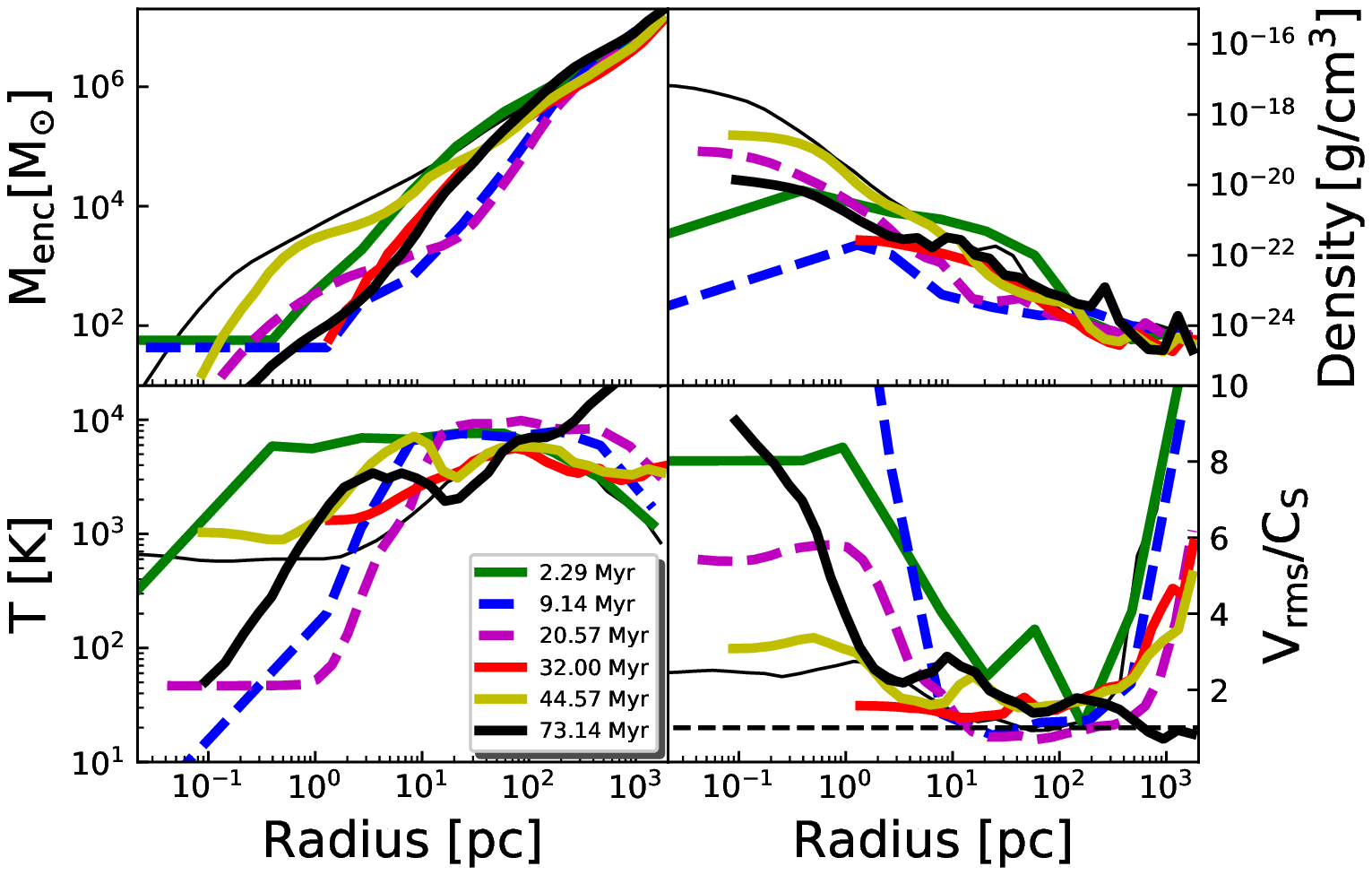}
\end{minipage} 
\end{tabular}
\caption{ The time evolution of  the averaged radial profiles of density, temperature, enclosed gas mass and turbulent velocity for halo 1 (top panel) and halo 2 (bottom panel) after the formation of a Pop III star is shown here. The thin solid line shows the state of the simulation at the onset of SF.  After 2.2 Myr a Pop III star dies a PISN and Pop II star cluster begin to form within a few Myr. They evacuate the gas from the halo center, drive outflows and turbulence.}
\label{fig3}
\end{figure*}

In halos 1 and 2,  after  the death of a Pop III star as a PISN,  dense cold clumps are formed by  metal cooling in the SN ejecta.  They result in a starburst  within 2 Myrs  leading to the formation of Pop II star clusters of   $\rm 954~M_{\odot}$ and $\rm 450~M_{\odot}$  in halos 1 and  2, respectively. The radiative feedback from Pop II stars heats the gas in their vicinity and dissociates the cold dense gas. After 4 Myr, feedback from Pop II SNe further heats and evacuates the gas  from the halo center. Consequently, the density drops below $\rm 10^{-24} ~g/cm^{3}$ in the central 10 pc, the temperature increases to a few thousand K and cold dense  gas gets depleted.  Star formation remains halted  for a few Myr and even up to about 20  Myr at one occasion.  In Fig. \ref{fig3} we show  profiles of density, enclosed gas mass, temperature and turbulent Mach number for halo 1 and halo 2, respectively.  The density varies from $\rm 10^{-24}-10^{-17} ~g/cm^{-}$ and bumps in the density profile indicate the presence of dense clumps. The temperature in the center cools down to 30 K  in the central 10 pc and is a few thousand K above  10 pc  due to the  longer cooling time at  lower densities of $\rm 10^{-24}-10^{-23} ~g/cm^{3}$. The time evolution of gas density and temperature shows the evacuation and heating of the gas in the halo center. The gas mass distribution in the central 10 pc is severely effected by  SN feedback and decreases by two orders of magnitude. In halo 1 the mass profile is recovered after 70 Myr but in halo 2 the gas mass in the central 10 pc is about an order of magnitude lower.  The typical turbulent Mach number in the central 10 pc is larger than 1 suggesting that turbulence is supersonic.

\begin{figure*}
\hspace{-6.0cm}
\centering
\begin{tabular}{c c}
\begin{minipage}{6cm}
\vspace{-0.2cm}
\hspace{-0.2cm}
\includegraphics[scale=0.54]{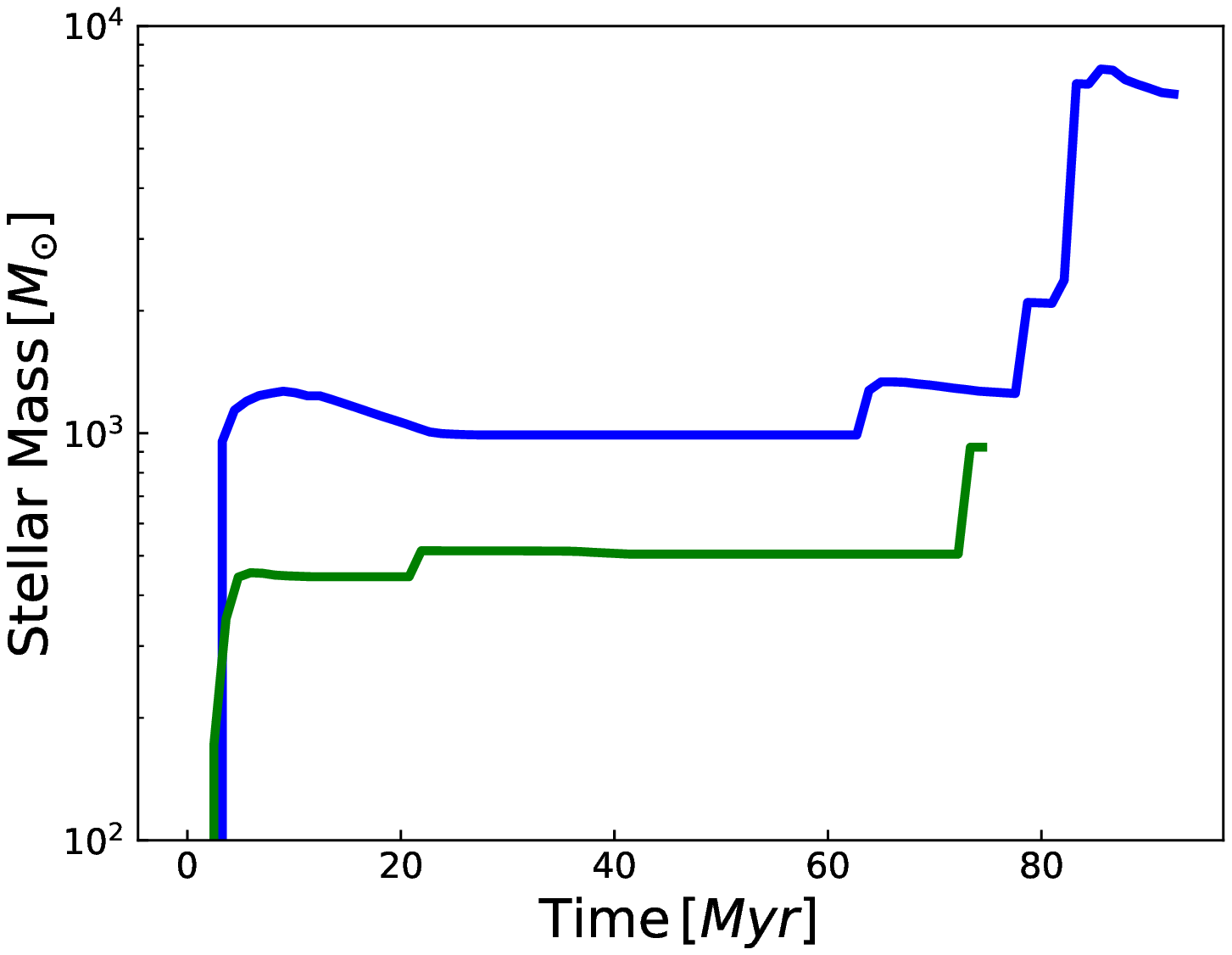}
\end{minipage} &
\begin{minipage}{6cm}
\hspace{1.8cm}
\includegraphics[scale=0.54]{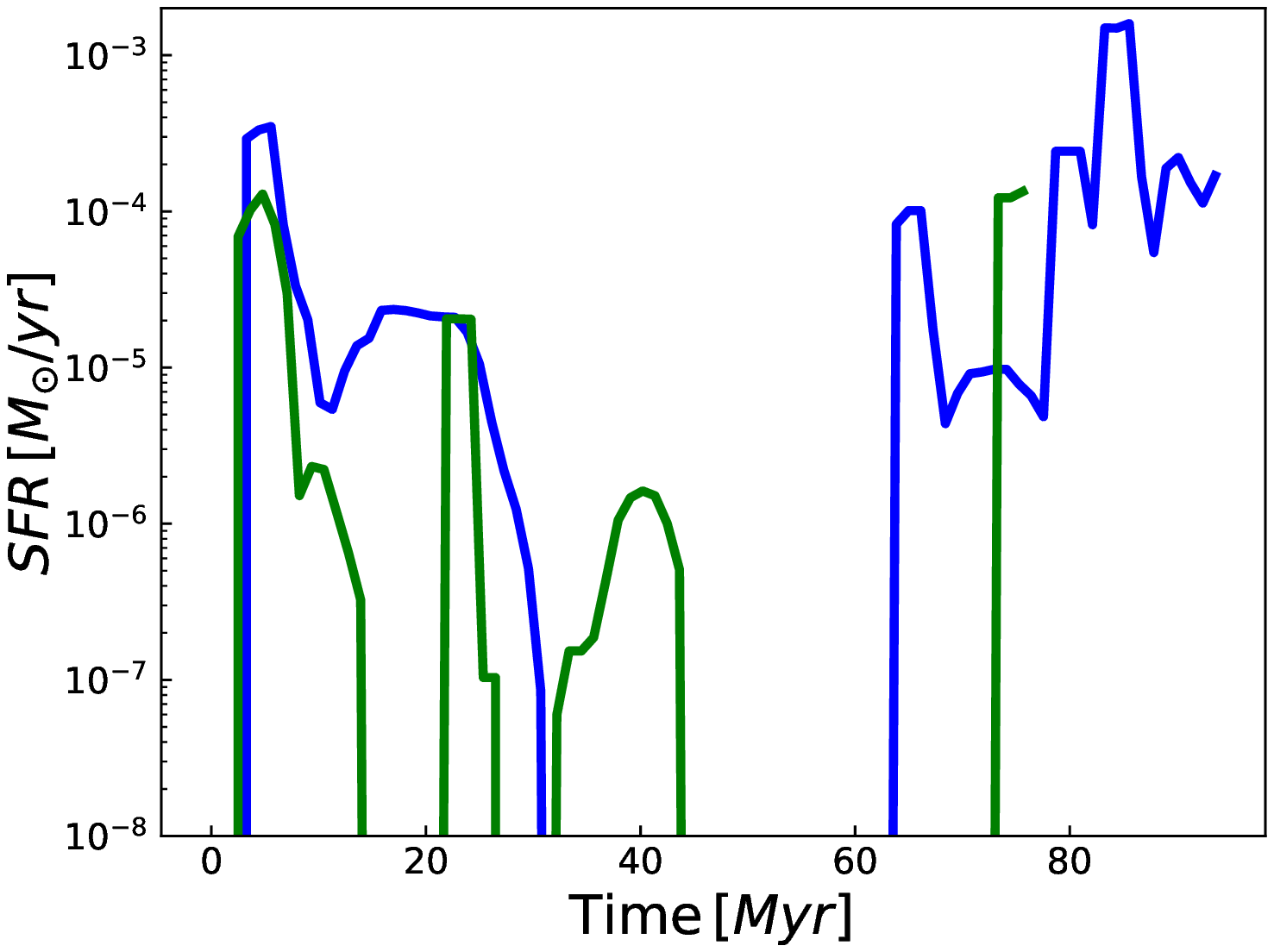}
\end{minipage} \\
\begin{minipage}{6cm}
\hspace{-0.2cm}
\includegraphics[scale=0.54]{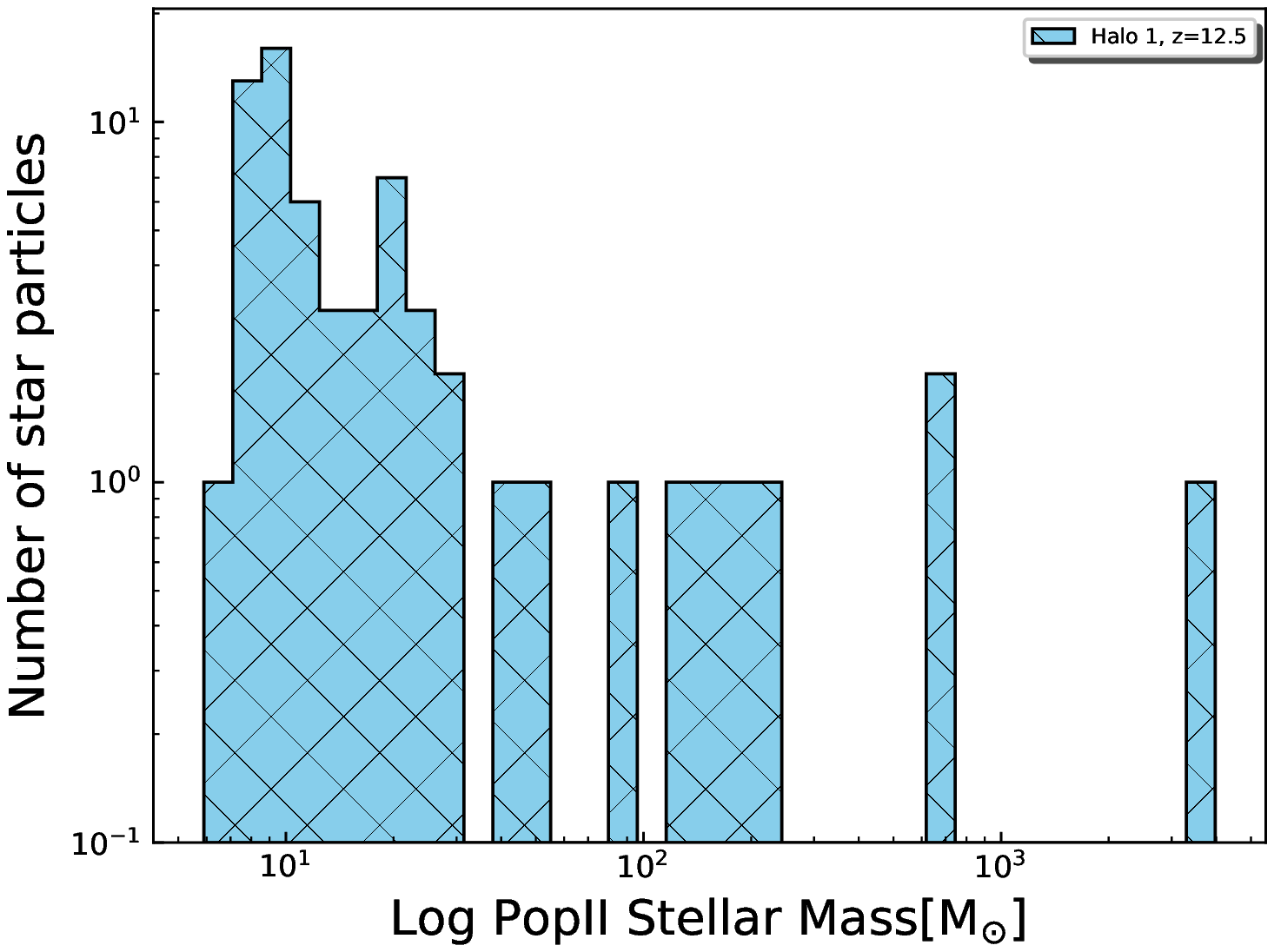}
\end{minipage} &
\begin{minipage}{6cm}
\hspace{2.0cm}
\includegraphics[scale=0.54]{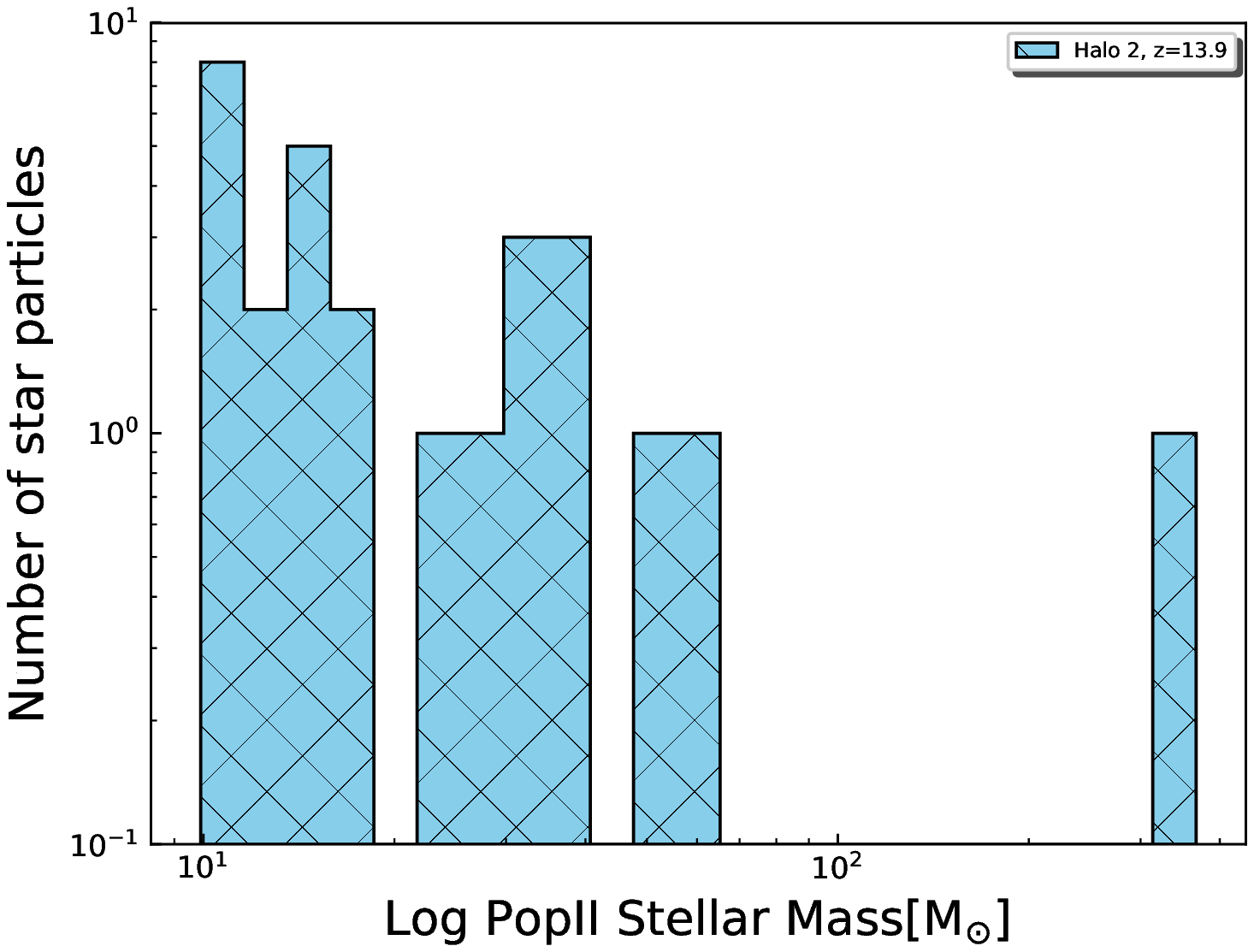}
\end{minipage} \\ 
\begin{minipage}{6cm}
\hspace{-0.2cm}
 \includegraphics[scale=0.54]{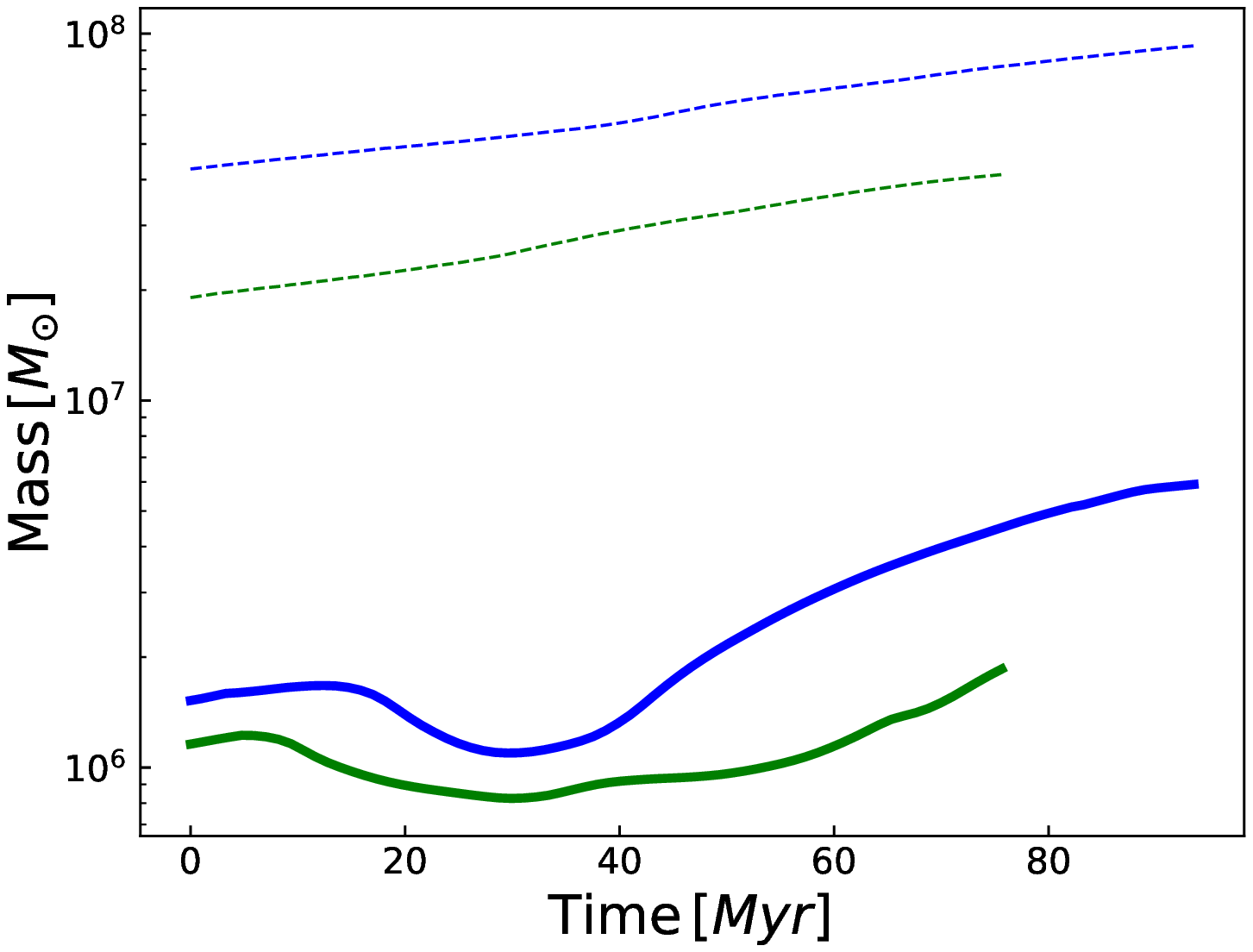}
\end{minipage} &
\begin{minipage}{6cm}
\hspace{2.50cm}
\includegraphics[scale=0.54]{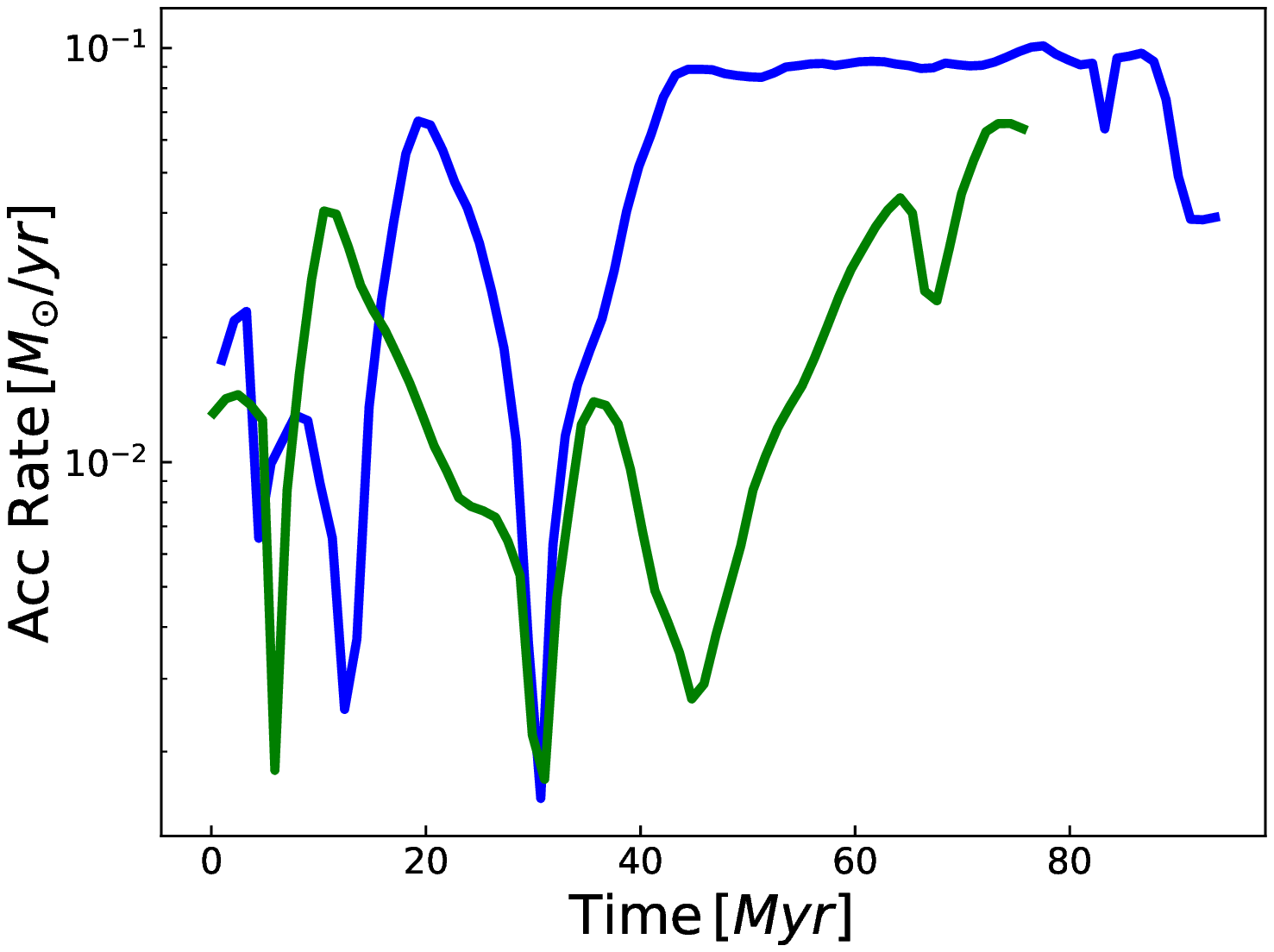}
\end{minipage} 
\end{tabular}
\caption{ The time evolution of stellar mass and SFR in halos 1 and 2 (top panel), the mass distribution of Pop II star particles (middle panel) and  gas mass and mass accretion rate into the central 300 pc (bottom panel). The blue line corresponds to halo 1, the green line halo 2 and the dashed lines correspond to the total halo mass.}
\label{fig8}
\end{figure*}

The time evolution of stellar mass and star formation rates (SFR)  for halo 1 and 2 is shown in Fig. \ref{fig8}. In halo 1, the initial increase in the stellar mass is due to the starburst in the aftermath of a PISN which results in $\rm \sim 1000~M_{\odot}$. Stellar mass continues to increase, reaches  $\rm  1236~M_{\odot}$ and a small decline is due to the mass  loss via SNe.  The stellar mass remains constant for the next 50 Myr as  SNe  evacuate the gas from the halo center and no significant SF occurs in halo 1 for 60 Myr. The jump in the stellar mass at 60 Myr  is due to the formation of  massive star particles.  At  78 \& 82 Myr massive star particles of  $\rm  849~M_{\odot}$ and $\rm  4857~M_{\odot}$ form in the cold dense clumps which boost the stellar mass to $\rm  \sim 7220~M_{\odot}$. They  lose $\rm 40 ~M_{\odot}$ and  $\rm 900 ~M_{\odot}$ due to SNe, respectively. The total stellar mass  in halo 1 by the end of simulation is $\rm  \sim 6800~M_{\odot}$. The star formation history shows that the mode of star formation is bursty and mainly regulated by SN feedback alongside significant contributions from radiative feedback.  There are epochs such as between 40-60 Myr when SF completely shuts down due to the depletion of cold star forming gas. In the last 20 Myr, the stellar mass  is increased by a factor of 4.  SFR  varies from $\rm 10^{-6}-10^{-3}~M_{\odot}/yr$ and the average SFR in halo 1 is about $\rm \sim 10^{-4}~M_{\odot}/yr$. 
Similar to halo 1, the initial starburst in halo 2 yields a stellar mass of $\rm  443~M_{\odot}$ which remains almost constant  until 70 Myr, 10 Myr longer than in halo 1 except for a  short starburst at 20 Myr.  No star formation activity is observed between 50-70 Myr due to the lack of cold gas supply regulated  by SNe.  At 75 Myr the increase in stellar mass to  $\rm  923~M_{\odot}$ is due to the formation of  star particles of $\rm 366~M_{\odot}$, $\rm 38~M_{\odot}$ and $\rm 14~M_{\odot}$.  The star formation history of halo 2 shows that SF occurs in short bursts and the SFR varies from $\rm 10^{-7}-10^{-4}~M_{\odot}/yr$. The mean SFR in halo 2 is $\rm  1.2 \times 10^{-5}~M_{\odot}/yr$. Overall,  halo 2 is more prone to SN feedback due to its shallower DM potential. Compared to halo 1, the average SFR in halo 2 is 10 times lower and the stellar mass  a factor of 7.3  smaller.

\begin{figure*} 
\begin{center}
\includegraphics[scale=0.8]{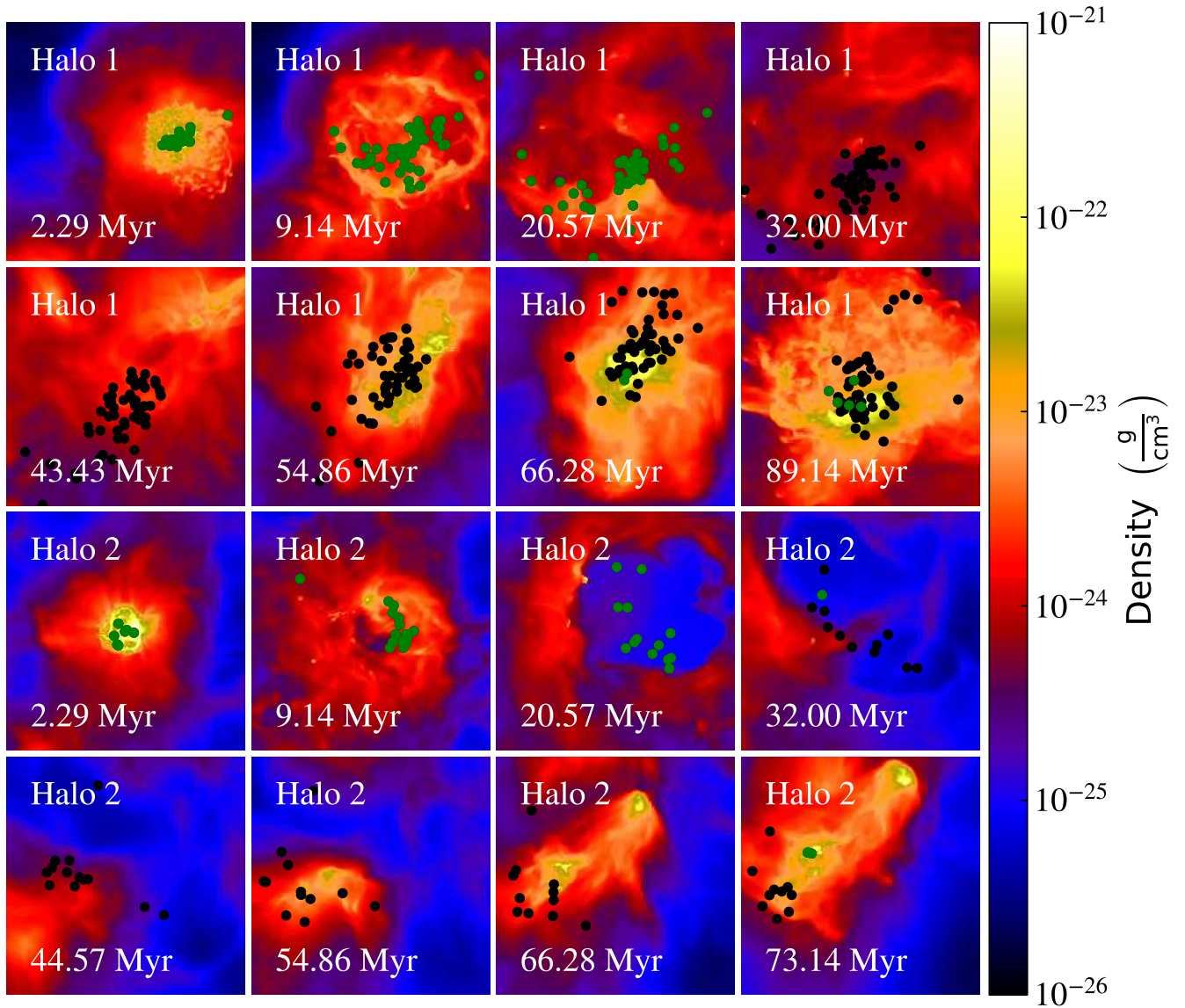}
\end{center}
\caption{ Time evolution of the density distribution in the halo 1 and halo 2 for the central 500 pc after the formation of a Pop III star. The green and black points represent young (ages $<$ 20 Myr) and old (ages $>$  20 Myr) Pop II stars. }
\vspace{-0.1cm}
\label{fig5}
\end{figure*}

\begin{figure*} 
\begin{center}
\includegraphics[scale=0.8]{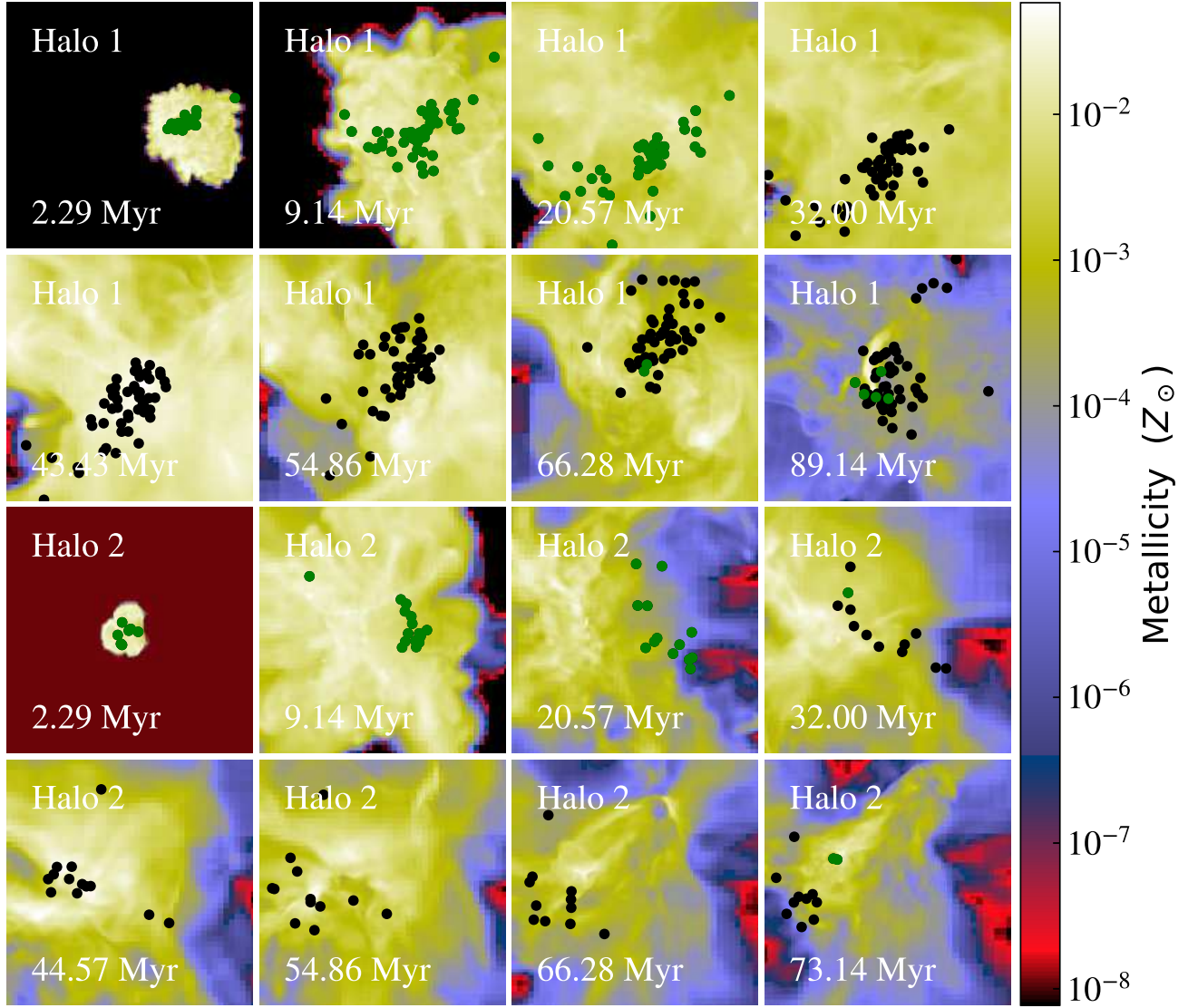}
\end{center}
\vspace{-0.1cm}
\caption{ Time evolution of the metallicity distribution in the central 500 pc, same as Fig. \ref{fig5} .}
\label{fig6}
\end{figure*}

The total gas mass in  halo 1 and halo 2 is $\rm \sim10^7~M_{\odot}$ and $\rm 5 \times 10^6 ~M_{\odot}$, respectively, but only a small fraction of  less than 1\% is turned into stars.  In figure \ref{fig5} we show the time evolution of the star particle distribution overplotted on gas density  in both halos.  At the onset of Pop II star formation, the density distribution is more concentrated but  the radiative and SNe feedback from stars expels the gas and redistributes it in the halo.  Consequently the mean gas density   decreases down to $\rm 10^{-25}~g/cm^{3}$ in the center and the fallback  leads to an enhanced SFR during the last 20 Myr in the case of halo 1. The average metallicity in both halos is shown in Fig. \ref{fig6}, initially the metallicity is mainly concentrated in the halo center and spreads through the halo due to the turbulent mixing over time. The average metallicity in both halos is $\rm \sim 10^{-2}~Z_{\odot}$. Overall the metal distribution is inhomogeneous but the metal fraction in the dense gas is above the critical value. Hence,  no Pop III stars form. The same trend is observed for both halos.
The mass distribution of star particles is shown in Fig. \ref{fig8}. In total, we have 66 star particles in halo 1, most of them are in the mass range of $\rm 5-30~M_{\odot}$,  6 are between 100-200 $\rm M_{\odot}$ and only one with mass of $\rm 3969~M_{\odot}$. In halo 2, there are  27 star particles in total, only one is  $\rm 366~M_{\odot}$ while most of them have masses between 10-40 $\rm M_{\odot}$. The massive star particles are formed in the cold dense clumps.  They represent a small cluster of stars instead of  a single massive star. 

We have evolved the simulations  for 92 Myr and 74 Myr  after the formation of the first Pop III star in halos 1 and 2, respectively.  The  gas masses at the end of the simulations in halos 1 and 2 are  $\rm 1.4 \times 10^7~M_{\odot}$ and  $\rm 5 \times 10^6~M_{\odot}$. The gas mass in halo 1 is doubled  in about 80 Myr while in halo 2 it is increased by a factor of 1.5. This suggests that the growth of halo 1 is faster than for halo 2. The  gas mass in the central 300 pc for halo 1 and halo 2 is $\rm 6 \times 10^6~M_{\odot}$ and $\rm 1.8 \times 10^6~M_{\odot}$. The mass accretion rate onto the central 300 pc (where most of the star formation occurs) during the first 40 Myr is  $\rm \sim 0.02~M_{\odot}/yr$ for both halos, see Fig. \ref{fig8}. For halo 1 it is increased up to $\rm \sim 0.1~M_{\odot}/yr$ while for the halo 2 it  is $\rm \sim 0.06~M_{\odot}/yr$. We also compared the accretion timescale ($\rm M_{gas}/M_{in}/yr$)  with the mass loss time ($\rm M_{gas}/M_{out}/yr$) at the virial radius of the halo and find that the accretion time is about a factor of 10 shorter than the mass loss time for halo 1 but for halo 2 they are comparable.  The mass depletion time  ($\rm M_{gas}/SFR$) for both halos is about 10 Gyr.  Assuming that 1\% of the total gas in the halo turns into stars over time, we expect these clusters to grow up to $\rm 10^5~M_{\odot}$ and $\rm 5 \times 10^4~M_{\odot}$.

\begin{figure*}
\hspace{-6.0cm}
\centering
\begin{tabular}{c c}
\begin{minipage}{6cm}
\vspace{-0.2cm}
\hspace{-0.5cm}
\includegraphics[scale=0.54]{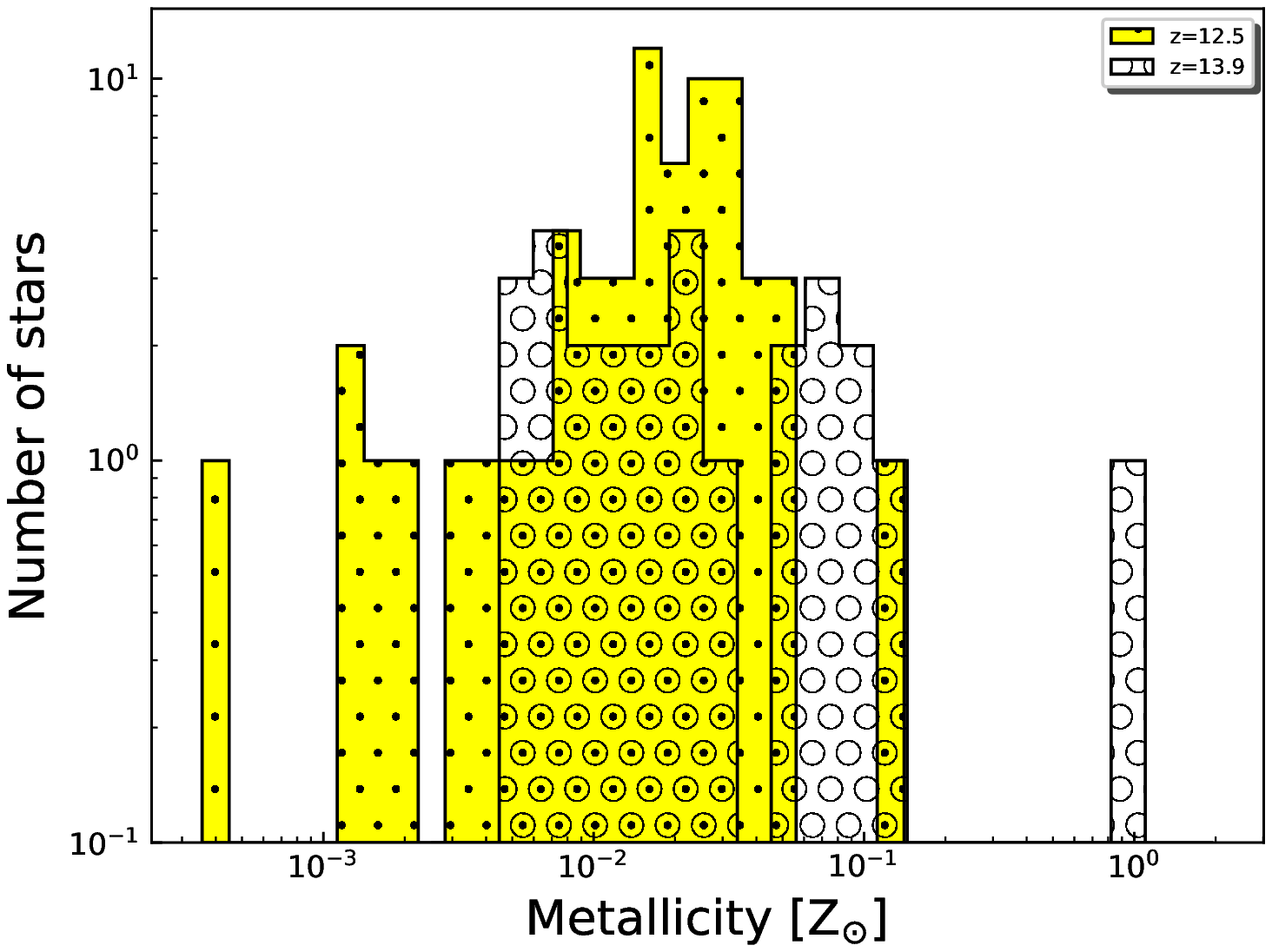}
\end{minipage} &
\begin{minipage}{6cm}
\hspace{2.50cm}
\includegraphics[scale=0.54]{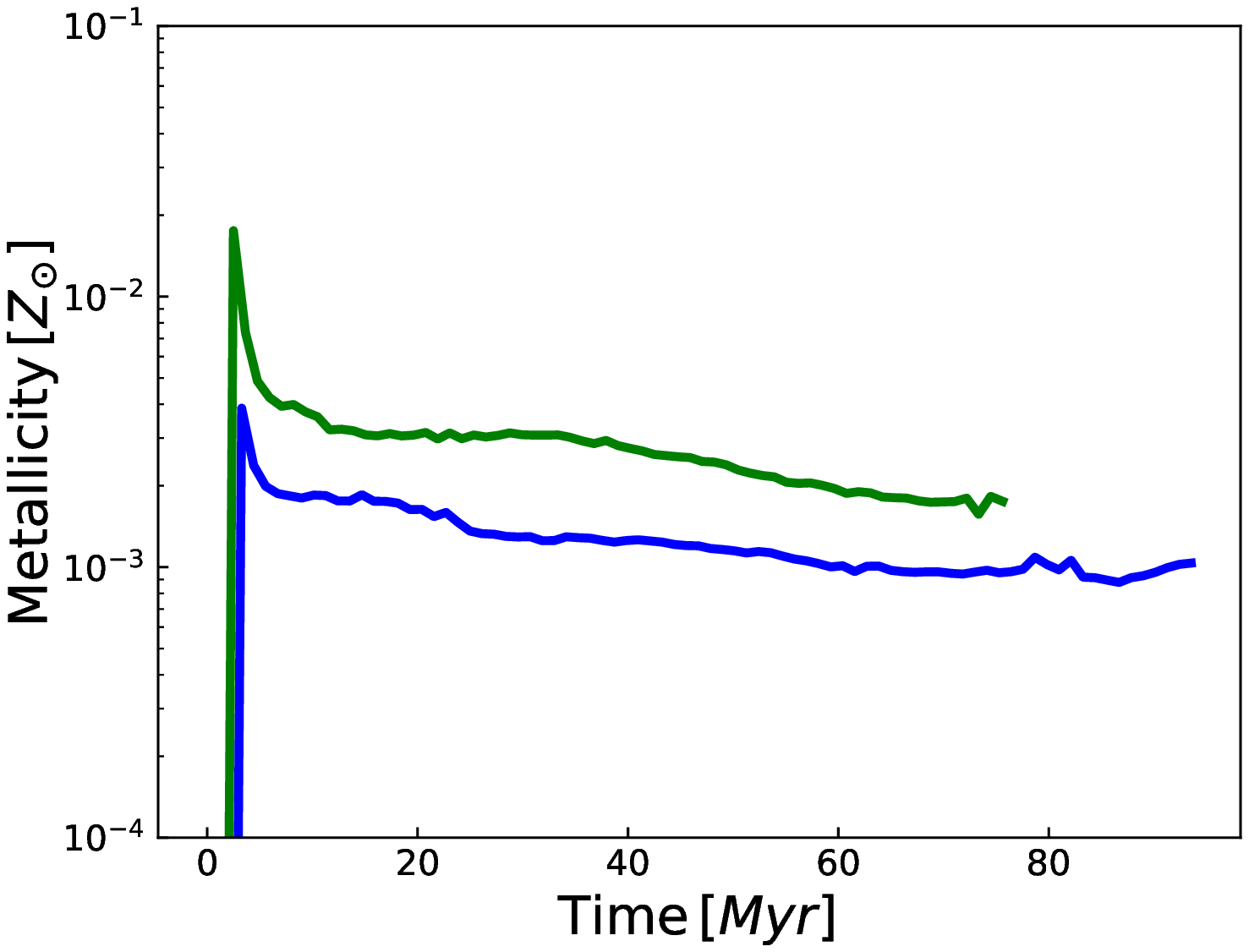}
\end{minipage} 

\end{tabular}
\caption{The metallicity distribution of Pop II star particles  (left) and the average metallicity of the halos (right).}
\label{fig7}
\end{figure*}

The  metallicity distribution of Pop II stars  ranges from $\rm  0.001-0.1~Z_{\odot}$ with an average value of a few times $\rm 0.01~Z_{\odot}$, see Fig. \ref{fig7}. This is an order of magnitude higher than the metallicity of  stars forming from type II SNe  \citep{Jeon15}. Our results show  that  the average metallicity in halos 1 and 2 is a few times  $\rm \geq 0.001~Z_{\odot}$ and  Pop II SF is suppressed in halos  of $\rm < 10 ^7~ M_{\odot}$ in agreement with previous studies \citep{Wise12,Muratov13,Jeon14}. Our estimates of SFRs and stellar masses are a factor of a few lower in comparison with the previous studies due to the energetics of a PISN \citep{Jeon15,Kimm16}.



\section{Conclusions} \label{sec:conc}


Our results show that a PISN  expels the gas from halos $\rm \leq 3  \times 10^{6}~M_{\odot}$  and shuts off  SF  for at-least 31 Myr. Halos  with mass $\rm > 10^7 ~M_{\odot}$ can retain gas and Pop II stars form in SN ejecta regulated by metal cooling. SF  occurs in  episodes  and is mainly regulated by Pop II SN feedback in tandem with radiative feedback. The mean SFR is  $\rm 10^{-5}-10^{-4}~M_{\odot}/yr$ and the star formation efficiency is $\rm \leq 1\% $.  Star clusters of $\rm  923~M_{\odot}$  and $\rm  \sim 6800~M_{\odot}$ form in halos of $\rm 3.8 \times 10^7~M_{\odot}$ and $\rm 9  \times 10^7~M_{\odot}$.  The average metallicity in the halos is  a few times $\rm \sim 10^{-3}~ Z_{\odot} $ well above the critical metallicity, consequently Pop III star formation shuts off in the  host halos.



%
%
\bibliographystyle{aasjournal}

\end{document}